%% file: main.tex
\documentclass[a4paper,11pt]{article}
\pdfoutput=1 

\usepackage{jcappub} 

\usepackage[T1]{fontenc} 
\usepackage{siunitx}

\usepackage{xcolor}
\usepackage{soul}
\usepackage{caption}
\usepackage{subcaption}

\title{The Canfranc Axion Detection Experiment (CADEx): Search for axions at 90 GHz with Kinetic Inductance Detectors}

\begin{document}
\author[a]{Beatriz Aja,}
\author[b]{Sergio Arguedas Cuendis,}
\author[c]{Ivan Arregui,}
\author[a]{Eduardo Artal,}
\author[d]{R. Belén Barreiro,}
\author[d]{Francisco J. Casas,}
\author[e]{Marina C. de Ory,}
\author[f]{Alejandro Díaz-Morcillo,}
\author[a]{Luisa de la Fuente,}
\author[g]{Juan Daniel Gallego,}
\author[f]{Jose María García-Barceló,}
\author[h]{Benito Gimeno,}
\author[e]{Alicia Gomez,}
\author[i]{Daniel Granados,}
\author[d]{Bradley J. Kavanagh,}
\author[c]{Miguel A. G. Laso,}
\author[c]{Txema Lopetegi,}
\author[f]{Antonio Jos\'e Lozano-Guerrero,}
\author[e]{Maria T. Magaz,}
\author[e,1]{Jes\'us Mart\'in-Pintado, \note{Corresponding author.}}
\author[d]{Enrique Martínez-González,}
\author[b,j]{Jordi Miralda-Escud\'e,}
\author[f]{Juan Monz\'o-Cabrera,}
\author[f]{Jose R. Navarro-Madrid,}
\author[k]{Ana B. Nuñez Chico,}
\author[a]{Juan Pablo Pascual,}
\author[k]{Jorge Pelegrin,}
\author[k]{Carlos Peña Garay,}
\author[e]{David Rodriguez,}
\author[d]{Juan M. Socuéllamos,}
\author[l]{Fernando Teberio,}
\author[c]{Jorge Teniente,}
\author[d]{Patricio Vielva,}
\author[d]{Iván Vila,}
\author[d]{Rocío Vilar,}
\author[e]{Enrique Villa,}


\affiliation[a]{Departamento de Ingenier\'ia de Comunicaciones, Universidad de Cantabria, Plaza de la Ciencia, 39005 Santander, Spain}
\affiliation[b]{Institut de Ci\`encies del Cosmos, Universitat de Barcelona, Mart\'i\ i Franqu\`es 1,
08028 Barcelona}
\affiliation[c]{Institute of Smart Cities and Dept. of Electrical, Electronic and Communications Engineering. Public University of Navarra. Campus de Arrosadia. 31006 Pamplona, Spain}
\affiliation[d]{Instituto de F\'isica de Cantabria (IFCA), CSIC-UC, Avenida de Los Castros s/n, 39005 Santander, Spain}
\affiliation[e]{Centro de Astrobiolog\'ia (CSIC – INTA), Torrej\'on de Ardoz, 28850 Madrid, Spain}
\affiliation[f]{Departamento de Tecnolog\'ias de la Informaci\'on y las Comunicaciones. Universidad Polit\'ecnica de Cartagena. 30302 Cartagena, Spain}
\affiliation[g]{Centro Astronómico de Yebes, Centro de Desarrollos Tecnológicos (CDT), Instituto Geogr\'afico Nacional (IGN), Guadalajara 19080, Spain}
\affiliation[h]{Instituto de Física Corpuscular (IFIC), CSIC-University of Valencia, 46980 Valencia, Spain}
\affiliation[i]{IMDEA Nanociencia, Cantoblanco, 28049 Madrid, Spain}
\affiliation[j]{Instituci\'o Catalana de Recerca i Estudis Avan\c cats, Barcelona}
\affiliation[k]{Laboratorio Subterráneo de Canfranc, 22880 Canfranc-Estación, Spain}
\affiliation[l]{Anteral S.L., 31006 Pamplona, Spain}

\emailAdd{ajab@unican.es}
\emailAdd{sarguedas@icc.ub.edu}
\emailAdd{ivan.arregui@unavarra.es}
\emailAdd{artale@unican.es}
\emailAdd{barreiro@ifca.unican.es}
\emailAdd{casas@ifca.unican.es}
\emailAdd{mcalero@cab.inta-csic.es}
\emailAdd{alejandro.diaz@upct.es}
\emailAdd{fuenterm@unican.es}
\emailAdd{jd.gallego@oan.es}
\emailAdd{benito.gimeno@uv.es}
\emailAdd{agomez@cab.inta-csic.es}
\emailAdd{daniel.granados@imdea.org}
\emailAdd{kavanagh@ifca.unican.es}
\emailAdd{mangel.gomez@unavarra.es}
\emailAdd{txema.lopetegi@unavarra.es}
\emailAdd{antonio.lozano@upct.es}
\emailAdd{mmagaz@cab.inta-csic.es}
\emailAdd{jmartin@cab.inta-csic.es}
\emailAdd{martinez@ifca.unican.es}
\emailAdd{miralda@icc.ub.edu}
\emailAdd{juan.monzo@upct.es}
\emailAdd{joser.navarro@edu.upct.es}
\emailAdd{anunez@lsc-canfranc.es}
\emailAdd{pascualp@unican.es}
\emailAdd{cpenya@lsc-canfranc.es}
\emailAdd{jpelegrin@lsc-canfranc.es}
\emailAdd{drodriguez@cab.inta-csic.es}
\emailAdd{socuellamos@ifca.unican.es}
\emailAdd{fteberio@anteral.com}
\emailAdd{jorge.teniente@unavarra.es}
\emailAdd{vielva@ifca.unican.es}
\emailAdd{vila@ifca.unican.es}
\emailAdd{evilla@cab.inta-csic.es}

\abstract{We propose a novel experiment, the Canfranc Axion Detection Experiment (CADEx), to probe dark matter axions with masses in the range \SIrange[range-phrase=--, range-units=single]{330}{460}{\micro \eV}, within the W-band (\SIrange[range-phrase=--, range-units=single]{80}{110}{\giga \hertz}), an unexplored parameter space in the well-motivated dark matter window of Quantum ChromoDynamics (QCD) axions. The experimental design consists of a microwave resonant cavity haloscope in a high static magnetic field coupled to a highly sensitive detecting system based on Kinetic Inductance Detectors via optimized quasi-optics (horns and mirrors). The experiment is in preparation and will be installed in the dilution refrigerator of the Canfranc Underground Laboratory. Sensitivity forecasts for axion detection with CADEx, together with the potential of the experiment to search for dark photons, are presented.
}

\maketitle
\flushbottom

\input{sec1}

\input{sec2}

\input{sec3}

\input{sec4}

\input{sec5}

\input{sec6}

\input{sec7}

\input{sec8}


\acknowledgments

We thank RADES team members for inspiring discussions. 
SA and JM are supported by grants PID2019-108122GB-C32 and
the Maria de Maeztu grant CEX-2019-000918-M of
ICCUB.
The work of UPCT and IFIC is supported by grant PID2019-108122GB-C33, funded by MCIN/AEI/10.13039/501100011033/ and by "ERDF A way of making Europe". JMGB thanks the grant FPI BES-2017-079787, funded by MCIN/AEI/10.13039/501100011033 and by "ESF Investing in your future". 
The work of Universidad de Cantabria is supported by the Ministry of Science and Innovation under Grant PID2019-110610RB-C22.
CAB and IMDEA-Nanoscience work is supported by grants PID2019-105552RB-C41 and PID2019-105552RB-C44 and by Comunidad de Madrid under Grant P2018/NMT-4291. IMDEA-Nanoscience acknowledges financial support from “Severo Ochoa” Programme for Centers of Excellence in R\&D (MINECO, Grant SEV-2016-0686). D.G. and A.G also acknowledge Grant DEFROST N62909-19-1-2053 from ONR Global. 
RBB, FJC, BJK, EMG, JMS and PV thank the Spanish Agencia Estatal de Investigaci\'on (AEI, MICIU) for the support to the Unidad de Excelencia Mar\'ia de Maeztu Instituto de F\'isica de Cantabria, ref. MDM-2017-0765. RBB, FJC, EMG and PV thank the Spanish Agencia Estatal de Investigaci\'on (AEI, MCI) for the funds received through the research project, ref. PID2019-110610RB-C21. RBB, FJC, BJK, EMG, JMS and PV also thank the `Dark Collaboration at IFCA' working group for useful discussions.
The work done by ANTERAL S.L. is supported by project QON-Space financed by the Navarra Government Project No. 0011-1365-2021-000220.
UPNA acknowledges financial support from the Spanish State Research Agency, Project No. PID2019-109984RB-C43/AEI/10.13039/501100011033 and Project No. PID2020-112545RB-C53/MCIN/AEI/ 10.13039/501100011033.

\bibliographystyle{JHEP}
\renewcommand{\refname}{References}
\bibliography{references.bib}

\end{document}

%% file: sec1.tex
\section{Introduction}
\label{sec:intro}
Cosmological and astrophysical observations of large-scale structure and the Cosmic Microwave Background Radiation~\cite{Bertone:2004pz,Clowe:2006eq,Planck2018} have demonstrated the existence of dark matter from its gravitational influence on baryonic matter and radiation. However, it has so far not been otherwise detected. Moreover, independent determinations of the dark matter and baryon cosmic densities based on the measured power spectrum for radiation and matter, the distance-redshift relation, direct mass measurements in galaxy clusters, and the light element abundances from nucleosynthesis, consistently indicate that dark matter constitutes 84.3\% of all matter in the universe, with the remaining 15.7\% being the known baryonic matter. In spite of this, the nature of dark matter remains a mystery and a key question for particle physics and cosmology. 

A particularly attractive dark matter candidate is the Quantum Chromodynamics (QCD) axion: it arises from a theory that solves a fundamental problem in the Standard Model (SM) of particle physics, the strong Charge Conjugation-Parity (CP) problem~\cite{Peccei:2006as}, and at the same time predicts the existence of Cold Dark Matter, which satisfies all present observational constraints. Due to the repeated null results of many worldwide efforts to detect weakly interacting massive particles (WIMPs)~\cite{Billard:2021uyg}, which have been the favorite dark matter candidate for the last three decades, the axion dark matter hypothesis has recently attracted increased interest from the experimental community.

The CP problem is the absence of CP violation in the strong force. The expected CP violation arises from the term of the SM Lagrangian for strong interactions containing the $\theta$ angle. This term induces an electric dipole moment for the neutron, which has an experimental upper limit of $\sim \SI{e-26}{e\cdot\centi \metre }$~\cite{Abel:2020gbr}, implying $\theta\lesssim 10^{-10}$. There is no reason for this $\theta$ angle to be so small compared to unity. 

The most elegant solution to the CP problem, the Peccei-Quinn (PQ) mechanism~\cite{Peccei:1977hh,Peccei:1977ur}, introduces a global U(1)$_\mathrm{PQ}$ symmetry and promotes $\theta$ to a dynamic field $\theta + a(x)/f_a$, where $\theta_i$ = $a/f_a$ is the misalignment angle, $a(x)$ is the axion field and $f_a$ is the axion scale. At energies below $f_a$ this PQ symmetry is spontaneously broken, generating the pseudo-scalar Goldstone boson known as the axion~\cite{Wilczek:1977pj, Weinberg:1977ma}.

Axions can be produced by the vacuum realignment mechanism~\cite{Visinelli:2009zm,Visinelli:2009kt}, with their abundance determined by $\theta_i$, and they acquire a mass because the U(1)$_\mathrm{PQ}$ symmetry is explicitly broken by the chiral anomaly. The mass of the axion is inversely proportional to the axion scale $f_a$. Two of the most popular benchmark axion models, the Kim~\cite{Kim:1979if}, and Shifman, Vainsthein, and Zakharov~\cite{Shifman:1979if} (KSVZ) model and the Dine, Fischler, Srednicki~\cite{Dine:1981rt} and Zhitnitsky~\cite{Zhitnitsky:1980tq} (DFSZ) model, postulate an $f_a$ significantly above the electroweak scale, consequently producing a very light axion (with mass between \SI{}{\micro \eV} and \SI{}{\eV}).

After the Peccei-Quinn model had been proposed to solve the strong CP problem, the axion was found to be an excellent Cold Dark Matter candidate because of its production mechanism and properties~\cite{Preskill:1982cy,Abbott:1982af,Dine:1982ah}. The calculation of the axion relic density ($\Omega_a$) and mass ($m_a$) depends on the detailed dynamics of the axion field in the presence of complicated finite-temperature QCD effects, resulting in an extensive range of possible masses. Moreover, $\Omega_a$ and $m_a$ depend on the cosmic epoch (post or pre-inflation) when the PQ symmetry is broken.

In the post-inflationary scenario, the PQ symmetry is broken after the end of inflation, and the initial misalignment angle $\theta_i$ takes on a different value in different causally disconnected regions of the Universe~\cite{Marsh:2015xka}. Early estimates pointed towards an axion mass range of \SIrange[range-phrase=--, range-units=single]{4}{8}{\micro \eV} required to account for the entire DM abundance $\Omega_a = \Omega_\mathrm{DM}$~\cite{Bonati:2015vqz,Petreczky:2016vrs}.
Subsequent work recognised the importance of the production and decay of topological defects such as strings in populating the Universe with DM axions. Simulations accurately resolving these effects remain challenging, but recent estimates point to an axion mass greater than \SI{20}{\micro \eV}, with viable masses up to $\sim \SI{500}{\micro \eV}$~\cite{Berkowitz:2015aua,Fleury:2015aca,Ballesteros:2016euj, Borsanyi:2016ksw, Dine:2017swf,Klaer:2017ond, Buschmann:2019icd, Buschmann:2021sdq}.
The axion mass range $m_a \gtrsim \SI{40}{\micro \eV}$ remains largely unexplored by axion experiments, motivating new experimental searches for heavier DM axions.

The inverse Primakoff effect~\cite{Pirmakoff:1951pj,Sikivie:1983ip}, which converts axions into photons in the presence of a magnetic field, is one method of searching for the axion. The haloscope approach, developed by Sikivie in 1983~\cite{Sikivie:1983ip}, searches for axions in the local Galactic dark matter halo using a resonant microwave cavity within a massive superconducting magnet. A haloscope transforms axions to photons in a spectral line with a central frequency dictated by the axion mass and a line width determined by the kinetic energy of the axion, proportional to the squared velocity dispersion of the Milky Way dark matter halo in the solar vicinity. Because the mass of the hypothetical axion is unknown, experiments must search a wide frequency range for this spectral line. The figure of merit $F$ to improve the sensitivity of the detection experiment for a given axion-photon coupling $g_{a\gamma}$ is proportional to~\cite{PhysRevD.64.092003}:
 \begin{equation}
\label{eq:Figure-of-Merit}
    F \sim g_{a\gamma}^2 m_{a}^{-1}B^2V T_{\text{sys}}^{-1} C Q_\ell \,,
\end{equation}
where $B$ is the magnetic field strength; $V$ is the volume of the resonant cavities; $C$ is the form factor of the cavity (determined by the overlap between the dynamic electric field generated in the resonant cavity and the static magnetic field); $Q_\ell$ is the loaded quality factor of the cavity described in Sec.~\ref{sec:coupling_system}; $m_a$ is the axion mass; and $T_{\text{sys}}$ is the system noise temperature which is related to the Noise Equivalent Power (NEP) of the detectors as explained in section~\ref{sec:tech}.

\begin{figure}[ht]
    \centering
\includegraphics[width=0.9 \textwidth]{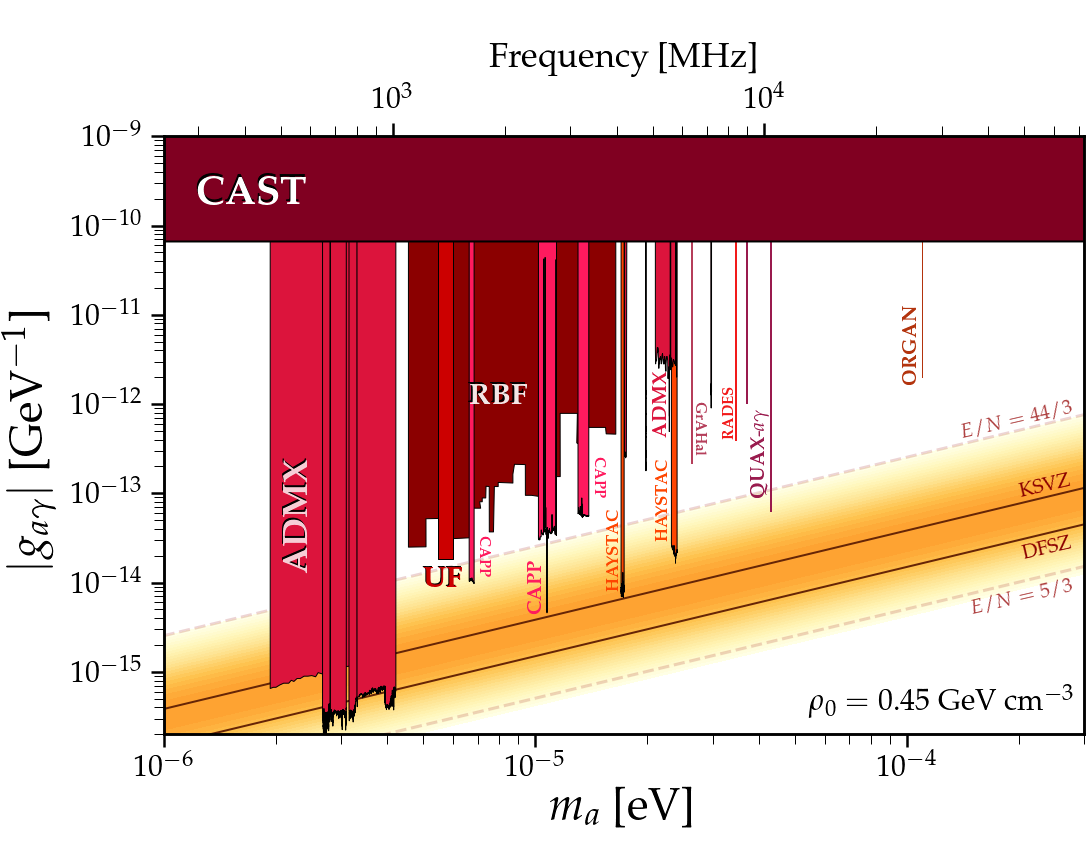}    
\caption{Current limits of the axion-photon coupling provided by the experiments using the haloscope technique and the CAST helioscope ~\cite{CAST:2017uph}. The two diagonal brown lines, the orange and yellow region represent the theoretical values given by the axion benchmark models. Taken from~\cite{ohare}.}
    \label{fig:AxionSearchSpace}
\end{figure}

Several experiments such as RADES~\cite{CAST:2020rlf}, CAPP~\cite{Jeong:2017hqs,Choi:2020wyr}, ADMX~\cite{Du:2018uak,Braine:2019fqb}, ADMX Side-Car~\cite{Boutan:2018uoc}, HAYSTAC~\cite{Zhong:2018rsr,Backes:2020ajv}, QUAX~\cite{Alesini:2019ajt, Alesini:2020vny}, ORGAN~\cite{McAllister:2017lkb}, GrAHal~\cite{Grenet:2021vbb} and others have used this technique to search for axions at frequencies between \SI{400}{\mega \hertz} and \SI{12}{\giga \hertz} (\SIrange[range-phrase=--, range-units=single]{1.65}{49.6}{\micro \eV})\footnote{The mass frequency relationship is given by $\nu_a = m_ac^2/h$} (see figure \ref{fig:AxionSearchSpace}). Above \SI{12}{\giga \hertz}, the axion parameter space is heavily unexplored, despite the fact that cosmological theory suggests that the dark matter axion may well be at higher mass. The main reasons are the difficulty in scaling the haloscope technique to higher frequencies (where the smaller resonant cavities imply a smaller detection volume), and the standard quantum limit to the sensitivity of heterodyne detectors. Beyond this, a number of novel detector concepts, considering broadband haloscopes and 
single-photon detection systems with potential sensitivities unlimited by the standard quantum limit, have been proposed to cover a wide range of axion masses. These include tunable plasma haloscopes (e.g.~ALPHA~\cite{Lawson:2019brd}), dielectric haloscopes~\cite{Millar:2016cjp} (e.g.~MADMAX~\cite{Beurthey:2020yuq}) and dish-antenna haloscopes~\cite{Horns:2012jf} (e.g.~BREAD~\cite{BREAD:2021tpx}). Recent reviews of experimental axion searches can be found in~\cite{Irastorza:2018dyq,Semertzidis:2021rxs,Sikivie:2020zpn}.

The haloscope setup for axion detection is also sensitive to other light, weakly coupled particles, including the dark photon $\gamma^\prime$. Dark photons (also known as hidden photons) are vector particles, kinetically mixed with the Standard Model photon~\cite{Fabbrichesi:2020wbt}. This coupling to electromagnetism induces an electric field in haloscope experiments, providing sensitivity to dark photon dark matter with a mass $m_{\gamma^\prime}$ which matches the resonant frequency of the haloscope (as in the case of the axion)
,without any dependence on the presence of a static magnetic field (unlike the axion). Crucially, such a search can typically be performed using the same data as an axion search or even using calibration data in the absence of a magnetic field~\cite{Ghosh:2021ard}.
Depending on the polarization state of the dark photon, it may also give rise to a time-varying signal due to the Earth's rotation~\cite{Caputo:2021eaa}. Through a careful choice of the observing schedule, it may be possible to detect this time-variation and therefore detect the dark photon. These considerations mean that a dark photon search can typically be performed with little additional experimental exposure time. Dark photons with masses around $m_{\gamma^\prime} \sim \SI{400}{\micro \eV}$ correspond to a region of parameter space where bounds are comparatively weak. In this range, the dark photon can be a viable Dark Matter candidate, produced for example through a realignment mechanism analogous to that of the axion~\cite{Nelson:2011sf,Arias:2012az,Alonso-Alvarez:2019ixv}. An experimental search for dark photons with sensitivity to kinetic mixing at the level of $\chi \sim 10^{-9} - 10^{-8}$ would therefore probe new, unconstrained parameter space~\cite{Caputo:2021eaa}. 

In this work, we propose a novel experiment, the Canfranc Axion Detection Experiment (CADEx), to search for the Dark Matter axion in the mass range (\SIrange[range-phrase=--, range-units=single]{330}{460}{\micro \eV}) within the  W-band (\SIrange[range-phrase=--, range-units=single]{80}{110}{\giga \hertz}). CADEx combines a microwave resonant cavity haloscope with a broadband incoherent detector system to be installed in the dilution refrigerator in the Canfranc Underground Lab (LSC)~\cite{lsc} in Spain, with the potential for also searching for dark photons.

The paper is organised as follows: in section \ref{sec:tech} we compare coherent and incoherent detection techniques. The CADEx concept is described in section \ref{sec:design} and the proposed design of the  haloscope,  optics and detector components are described in sections \ref{sec:halo},  \ref{sec:optics} and \ref{sec:detect}, respectively. Finally, sensitivity expectations for the axion-photon coupling and the dark photon kinetic mixing are presented in section \ref{sec:sens}, and conclusions and future prospects are discussed in section \ref{sec:con}.

%% file: sec2.tex
\section{Detection techniques. Coherent versus incoherent }
\label{sec:tech}
The axion signature is expected as a very narrow emission feature in the frequency domain, and heterodyne receivers are the classical detection systems used in all axion detection experiments at low frequencies (<\SI{50}{\giga \hertz}). Heterodyne receivers amplify and convert the input signal from the haloscope to a lower frequency band, while preserving the information of amplitude and phase. The advantage of this system is that the down-converted signal can be easily processed and digitized to obtain its spectrum (via real-time Fast Fourier Transform) with very high frequency resolution, spectroscopically resolving the radiation generated in the haloscope. The sensitivity floor for the noise temperature of a heterodyne receiver is imposed by the standard quantum noise limit for a coherent detector ($\approx \SI{2.2}{\kelvin}$ at \SI{90}{\giga \hertz}~\cite{caves1982quantum}). Very low noise cryogenic W-band heterodyne receivers are routinely used in radio astronomy reaching a state-of-the-art performance noise temperature of $\approx$ \SIrange[range-phrase=--, range-units=single]{25}{30}{\kelvin} ~\cite{tercero2021yebes,yagoubov2020wideband} when cooled to a physical temperature of \SI{4}{\kelvin}. Practical semiconductor-based heterodyne detectors are not expected to improve from present values (\SIrange[range-phrase=--, range-units=single]{25}{30}{\kelvin}) in the short-midterm.\\

On the other hand, incoherent detectors, such as those based on bolometers, transition edge sensors (TES), kinetic inductance detectors (KID) or quantum capacitance detectors are not affected by the standard quantum noise limit, as heterodyne receivers are~\cite{caves1982quantum}. Their sensitivity is characterized by the Noise Equivalent Power (NEP), defined as the minimum detectable power per square root bandwidth \SI{}{(\watt \per \sqrt{\hertz})}~\cite{leclercq_NEP}, which is limited by photon noise and other factors related to technology. These detectors use superconductor material properties and they can provide high sensitivities in the W-band and in higher frequencies. TES bolometers use a superconductor as a resistive thermometer, whereas the KID detection mechanism exploits the changes of the superconducting kinetic inductance caused by absorbed photons~\cite{ulbricht2021applications}. While TES bolometers operate at the superconducting transition temperature, $T_\text{c}$,  KIDs operate at temperatures well below $T_\text{c}$, with conduction electrons in the form of Cooper pairs and identically zero DC resistance. The lowest optical NEP demonstrated so far in a TES bolometer is \SI{3e-19}{\watt \per \sqrt{\hertz}}~\cite{TES21,karasik2011demonstration, nagler2021transition}, and KID technology has reached a NEP sensitivity of \SI{3.8e-19}{\watt \per \sqrt{\hertz}}~\cite{Visser2014fluctuations}.

To make a  direct comparison between  the sensitivity of the two systems in terms of the NEP, noise temperature and signal-to-noise ratios, we consider idealized coherent and incoherent detectors. Typically, the sensitivity of coherent receivers is described in terms of noise temperature, and the signal-to-noise ratio (SNR) is given by
\begin{equation}\label{SNR_coh}
    \text{SNR}_{\text{coh}} =\frac{T_\text{s}\sqrt{\tau~\Delta \nu }}{T_{\text{sys}}}\,,
\end{equation}
where $T_\text{s}$ is the brightness temperature of the signal, $T_{\text{sys}}$ is the system noise equivalent temperature (sum of the background temperature $T_{\text{bkg}}$ and receiver noise temperature $T_{\text{rec}}$), with a resolution bandwidth $\Delta \nu$, and integration time $\tau$. In this case, since the spectrum can be resolved, the maximum SNR is obtained by adjusting the resolution of the instrument to the bandwidth of the axion signal ($\Delta \nu = \Delta \nu_\text{s}$).
The SNR of an incoherent receiver is
\begin{equation}\label{SNR_inc}
    \text{SNR}_{\text{inc}} =\frac{P_\text{s}\sqrt{2\tau}}{\text{NEP}}\,,
\end{equation}
where $P_\text{s}$ is the signal power calculated as $k_BT_\text{s}\Delta \nu_\text{s}$ ($k_B$ is the Boltzmann constant and $\Delta \nu_\text{s}$ is the signal bandwidth).

The cavity resonant frequency of the haloscope must be tuned to the axion mass, within the haloscope bandwidth $\sim \nu/Q_l$ (where $Q_l$ is the cavity loaded quality factor, defined in section~\ref{sec:halo}), for the axion signal to be produced. The conversion of axions to photons is expected to produce a narrow emission peak of fractional width $\thicksim10^{-6}$, determined by the Galactic halo velocity dispersion, in the cavity power spectrum~\cite{hagmann1998results}. The cavity generates a peak of linearly polarized thermal noise of bandwidth $\nu/Q_l$, which would ideally be as narrow as the expected signal width but is typically much broader. The noise background, $T_{\text{bkg}}$, arises mainly from the haloscope physical temperature, which can be reduced to mK in our experiment. Therefore, the system noise temperature is dominated by the much higher receiver temperature for a heterodyne receiver, and the background power will also be subdominant for an incoherent detector compared to its typical NEP values.

Considering the case of detection at \SI{90}{\giga \hertz},  the expected axion signal bandwidth is \SI{90}{\kilo \hertz}. In this case,  a coherent receiver with a state-of-the-art $T_{\text{sys}}=\SI{25}{\kelvin}$ and an incoherent detector
with NEP $=$ \SI{1.46e-19}{\watt \per \sqrt{\hertz}}, (NEP$=k_BT_{\text{sys}}\sqrt{2\Delta \nu}$) will provide the same SNR. 
However, KID technology has the potential to exceed this sensitivity requirement, reaching a NEP around \SI{1e-20}{\watt \per \sqrt{ \hertz}} using different
strategies~\cite{hailey2021kinetic}. We therefore consider KIDs as the baseline detector technology for CADEx.

%% file: sec3.tex
\section{Conceptual design}
\label{sec:design}
CADEx will search for axions in the mass range \SIrange[range-phrase=--, range-units=single]{330}{460}{\micro \eV} within the W-band (\SIrange[range-phrase=--, range-units=single]{80}{110}{\giga \hertz}) by combining the haloscope approach with an incoherent detection system based on KID technology. Incoherent detectors are  broad band receivers which do not provide the spectral resolution to detect the narrow-frequency feature produced by the axion in the haloscope. The incoherent detectors in CADEx will measure the linearly polarized axion signal generated in the haloscope against the unpolarized background emission as a function of the resonant frequency of the haloscope.

\begin{figure}[ht]
    \centering
    \includegraphics[scale=0.5]{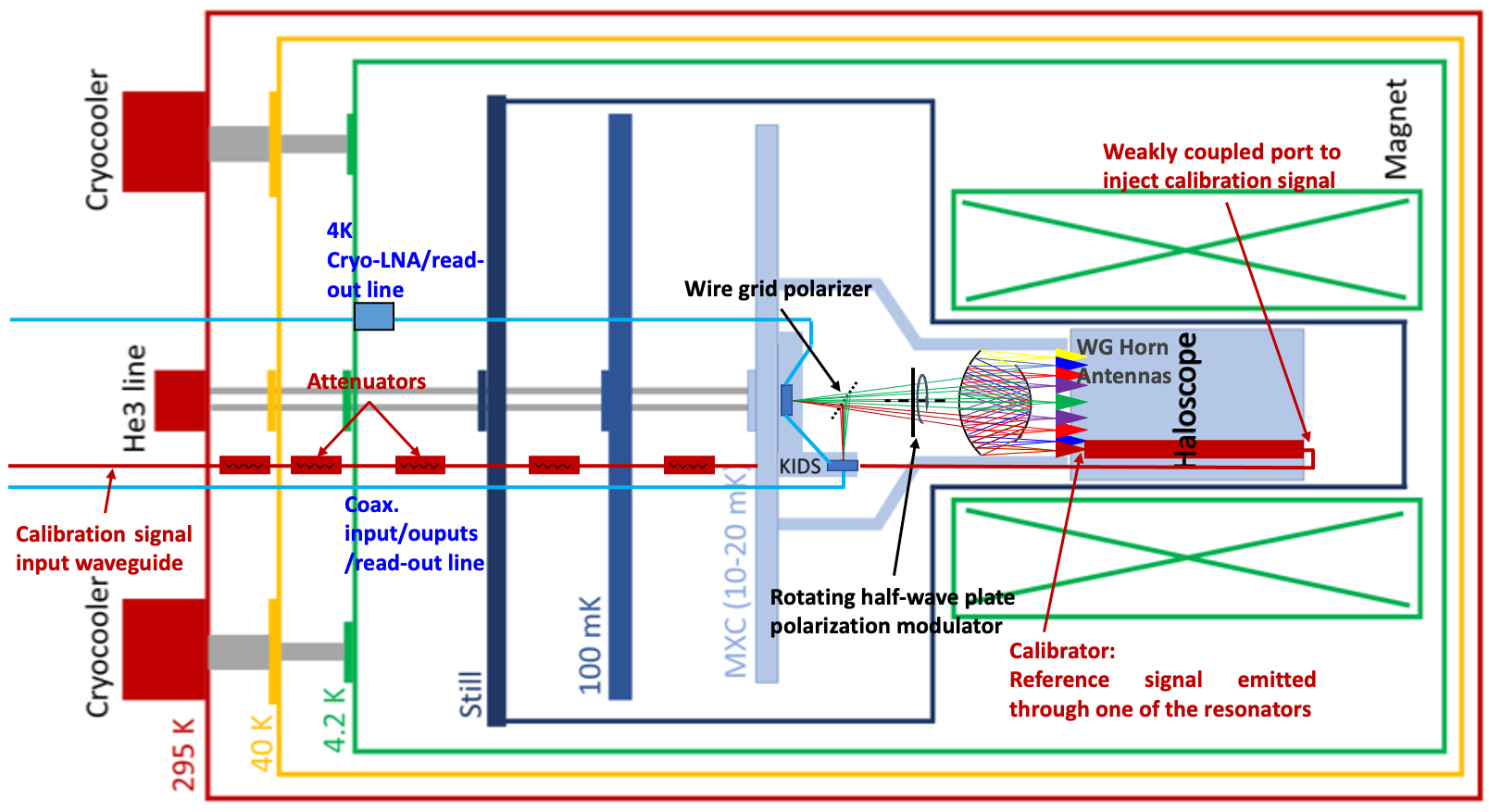}
    \caption{Schematic  block diagram proposed for the CADEx's accommodation in the dilution refrigerator of the Canfranc Underground Laboratory. The different temperature stages in the cryostat are indicated with different colors  (red: ambient, yellow: \SI{40}{\kelvin}, green: \SI{4.2}{\kelvin} and light blue: \SIrange[range-phrase=--, range-units=single]{10}{20}{\milli \kelvin}). The \SI{10}{\tesla} magnet operating at \SI{4}{\kelvin} is depicted by two green boxes with diagonal lines. The main CADEx subsystems installed in the mK stage are also shown: the haloscope (light blue inside the magnet, section \ref{sec:halo}), the optics (coloured horns and dash dotted rays, section \ref{sec:optics}) and the two KID arrays to measure two orthogonal linear polarizations (dark blue, section \ref{sec:detect}). The calibration signal injected externally through the different temperature stages is shown in red (section \ref{sec:optics}).}
    \label{fig:refrigerator}
\end{figure}

CADEx will be installed in the dilution refrigerator of the Canfranc Underground Lab (LSC) to decrease the impact of cosmic rays on the final sensitivity of the broadband incoherent detectors.  Figure \ref{fig:refrigerator} shows a block diagram of the experiment accommodated inside the LSC dilution refrigerator, indicating the location of the main subsystems and their temperature.  The microwave resonant cavity haloscope described in section \ref{sec:halo} will be located in the mK stage to minimize the background radiation seen by the KID detectors, in a static magnetic field of \SIrange[range-phrase=--, range-units=single]{8}{10}{\tesla}. The radiation from the haloscope will be combined through an optimized quasi-optics system with horns and mirrors (see section \ref{sec:optics}) and focused on the detection system.  As described in section \ref{sec:detect}, the detection system will make use of radio astronomy techniques \cite{NIKA2_aya} to measure the degree of linear polarization of the signal arising from the haloscope, tuned to two adjacent resonant frequencies (see section \ref{subsec:Tuning system}). The calibration of the system will be achieved by injecting a polarized signal of known intensity.

%% file: sec4.tex
\section{Haloscope design}\label{sec:halo} 
Following Sikivie's approach, the haloscope in the detection experiment will be a single cavity or a set of multiple resonant cavities working at the frequency of interest. The detected power from axion-photon conversion is then~\cite{PhysRevLett.51.1415,PhysRevLett.55.1797,AlKenany:2016trt}:
\begin{equation}
\label{Pd_eq}
    P_d = \frac{\beta}{(1+\beta)^2} g^2_{a\gamma}\frac{\rho_a}{m_a}B^2CVQ_0\,,
\end{equation}
where $\rho_a$ is the dark matter density, and a number of key parameters which depend exclusively on the haloscope design can be identified. These parameters are the cavity volume ($V$), the form factor ($C$) of the electromagnetic mode which couples with the axion-photon conversion, the coupling factor for the signal extraction ($\beta$), and the unloaded quality factor ($Q_0$). $V$, $C$ and $Q_0$ depend on the cavity geometry and the chosen electromagnetic mode, whereas $\beta$ depends additionally on the cavity coupling system. Therefore, the operational goals in the design of the haloscope, in order to maximize the detected power and maximizing the sensitivity to $g_{a\gamma}$, is to optimize the coupling $\beta$ while maximizing $V$, $C$ and $Q_0$.

Following the experience in RADES for \SI{8.4}{\giga \hertz} haloscopes~\cite{Diaz-Morcillo:2021psa}, we adopt rectangular geometries for the cavities and calculate the above parameters for this type of microwave resonators.

\subsection{Form factor}
\label{sec:form_factor}
The solenoid magnet in the LSC facility generates a static magnetic field that is constant and parallel to the magnet axis. Therefore, in order to maximize $C$, an electromagnetic mode with the electric field parallel to this magnetic field must be chosen. In a rectangular cavity, assuming the $z$-axis as the magnet axis, this is the TM$_{110}$, with $C=64/\pi^4 \approx 0.66$~\cite{Diaz-Morcillo:2021psa}.

\subsection{Quality factor}
The unloaded quality factor for a TM$_{110}$ in a rectangular cavity, assuming only conductor losses, is given by~\cite{balanis1989}
\begin{equation}
\label{Q0_eq}
    Q_{0\mathrm{TM}_{110}} = \frac{1}{2} \sqrt{\frac{\pi\sigma}{f_r\varepsilon}} \frac{d\left(a^2 + b^2\right)^{\frac{3}{2}}}{ab\left(a^2+b^2\right)+2d\left(a^3+b^3\right)} ~,
\end{equation}
\\ where $\sigma$ is the electrical conductivity of cavity walls, $f_r$ is the mode resonant frequency, $\varepsilon$ is the electrical permittivity inside the cavity, which will be normally the vacuum one, $\varepsilon_0$, and $a$, $b$ and $d$ are the width, height and length of the cavity, respectively. In order to improve this quality factor, full-copper cavities will be employed in CADEx.

\subsection{Volume}
The relationship between resonant frequency and rectangular cavity dimensions for both TE$_{mnp}$ and TM$_{mnp}$ modes is given by equation \ref{fr_eq}, where $c$ is the speed of light in free space, and $m$, $n$ and $p$ are the number of the sinusoidal variations of the electric field along the $x$, $y$ and $z$ axes, respectively:
\begin{equation}
\label{fr_eq}
    f_r = \frac{c}{2} \sqrt{\left(\frac{m}{a}\right)^2 + \left(\frac{n}{b}\right)^2 + \left(\frac{p}{d}\right)^2}\,.
\end{equation}
Taking this into account, detection setups for masses of the order of hundreds of \SI{}{\micro \eV} leads to very tiny cavities. As an example, a target resonant frequency of \SI{90}{\giga \hertz} for the TM$_{110}$ mode requires a side of \SI{2.35}{\milli \meter} in a cubic cavity, giving a volume of \SI{13}{\micro \liter}, far from the necessary volume for obtaining acceptable sensitivities. Therefore, since there is plenty of room in the LSC magnet bore, the challenge here is increasing the volume without decreasing the operation frequency of the haloscope. A combination of two different approaches is explored for CADEx.

\subsubsection{Large cavities}
\label{sec:large_cavities}
\begin{figure}[ht]
    \centering
    \includegraphics[scale=0.8]{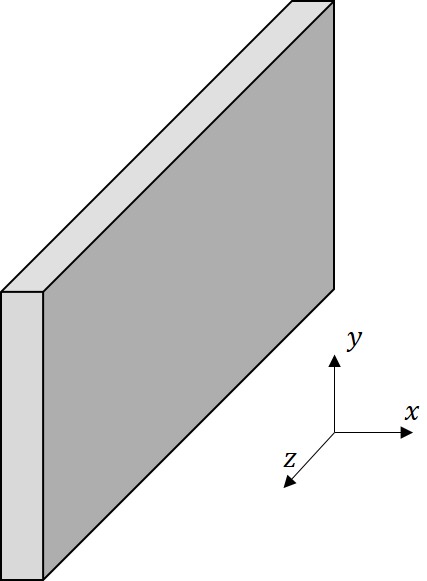}
    \caption{Large rectangular cavity, where the resonant frequency is determined mainly by its width (a).}
    \label{fig:large_cavs}
\end{figure}

\begin{figure}[ht]
    \centering
    \includegraphics[scale=0.5]{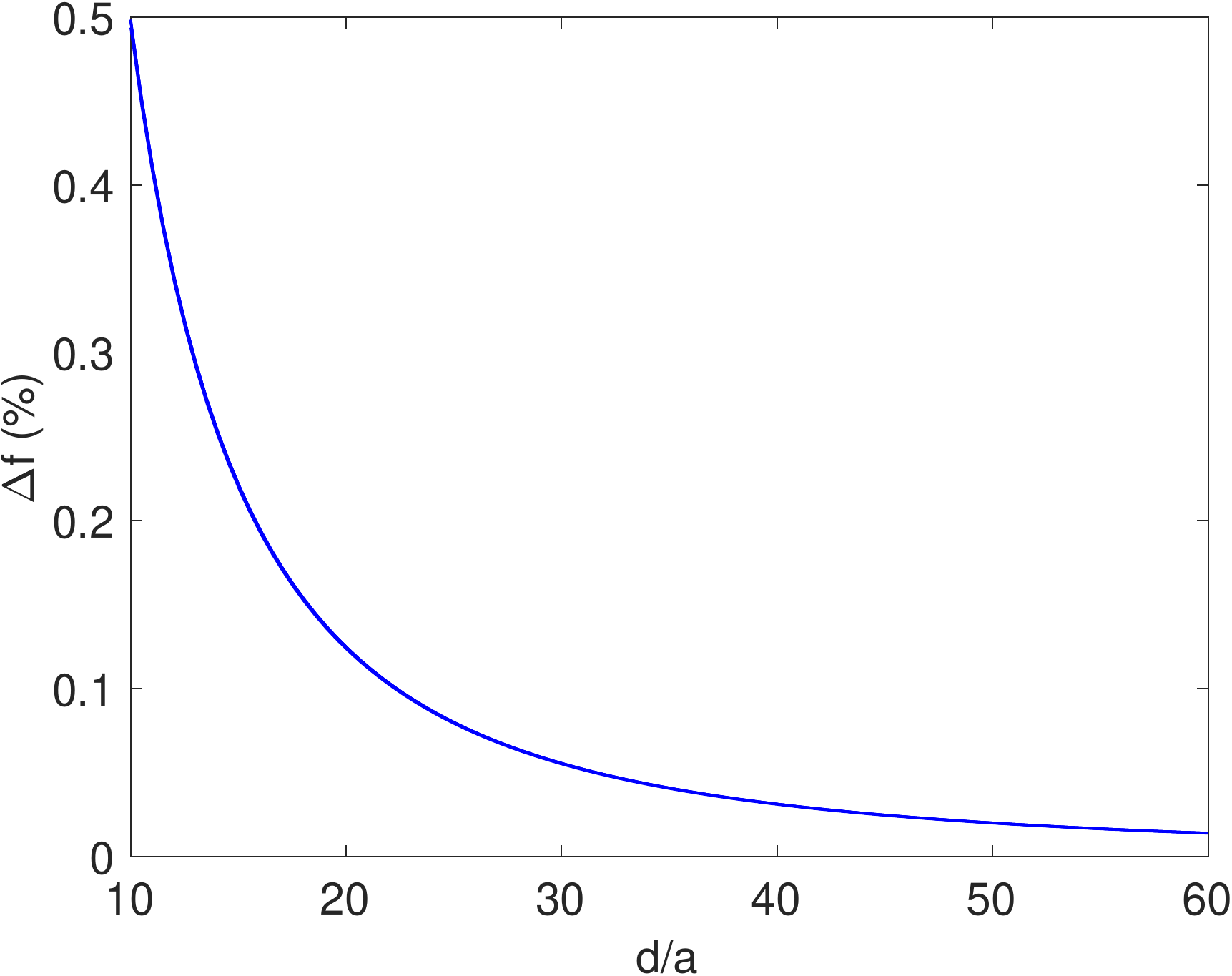}
    \caption{Relative frequency separation between modes TM$_{110}$ and TM$_{111}$ in a rectangular cavity when $d$ increases.}
    \label{fig:DeltafTM110TM111}
\end{figure}

For the mode TM$_{110}$, equation (\ref{fr_eq}) reduces to 
\begin{equation}
\label{frTM110_eq}
    f_r = \frac{c}{2} \sqrt{\frac{1}{a^2} + \frac{1}{b^2}}\,,
\end{equation}
which allows us to increase the cavity length ($d$) without modifying the resonant frequency. Moreover, this resonant frequency is mainly determined by the cavity width ($a$) when its height ($b$) is large enough. In that case, increasing the height  hardly changes the resonant frequency. Therefore, the volume can be increased with longer and taller cavities, as shown in figure~\ref{fig:large_cavs}. The limit in this enlargement comes from the clustering of modes near the operation mode, which can hinder the detection of the mode through a vector network analyzer (VNA) and even reduce the form factor when two modes are almost overlapped. This problem worsens when a range of frequencies is explored, since the number of mode crossings increases. Figure~\ref{fig:DeltafTM110TM111} shows the relative frequency separation between the axion mode and its closest neighbor (with the same polarization) with increasing lengths. A trade-off between mode separation and volume can be found for $b=40a$ and $d=60a$. In that case, a separation of \SI{12.5}{\mega \hertz} ($0.014\%$) and a volume of \SI{11.1}{\milli \liter} are obtained. \\

\subsubsection{Multiple cavities}
\begin{figure}[ht]
    \centering
    \includegraphics[scale=0.8]{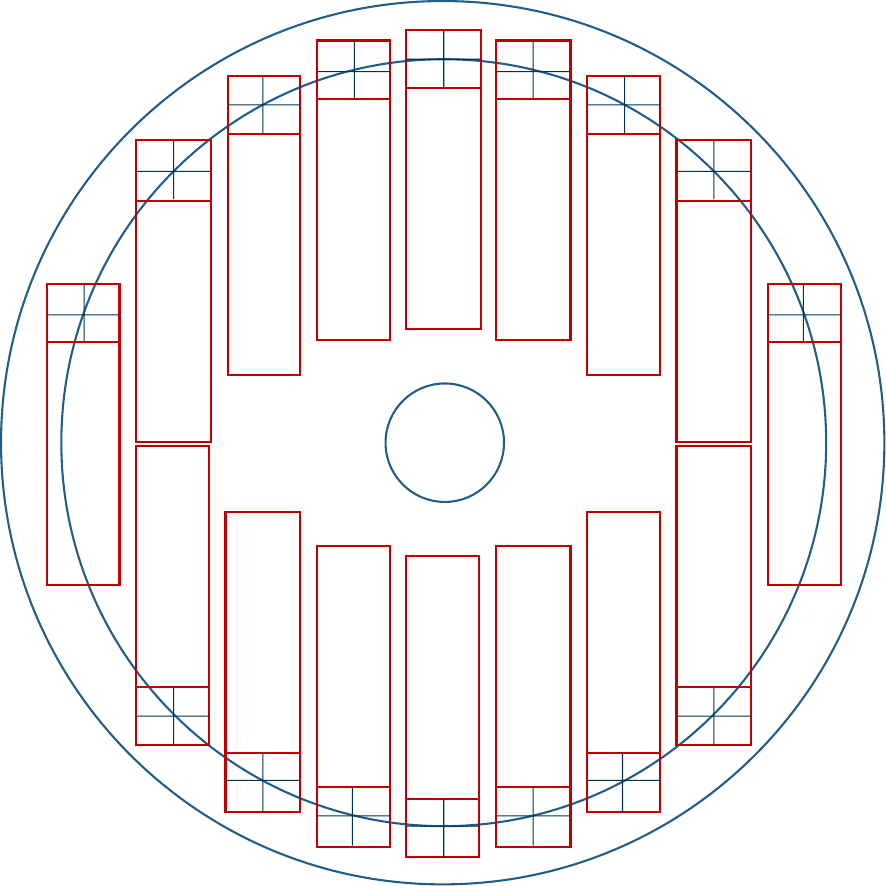}
    \caption{Circumference–arranged multiple cavities (16). Each cavity (red rectangles) is a large cavity (as described in section~\ref{sec:large_cavities}). The cross at each cavity points out the center of the horn antenna. These antennas are all pointing to a receiver antenna which is located in the center of the circumference, but displaced in the $z$ axis. The scheme is not to scale.}
    \label{fig:circumf_cavs}
\end{figure}

A larger volume can be obtained by means of the coherent sum of the signal extracted from $N$ resonant cavities, each one resonating at the same frequency. When this occurs, the total detected power is the sum of the powers from each individual cavity. This coherent sum requires the $N$ signals to be in phase at the combining device, in this case the receiving antenna. This is achieved when each signal travels the same electrical length from the cavity–line coupling point to the combining point. An example of such a configuration, with horn antennas pointing to a centered receiver antenna, is depicted in figure~\ref{fig:circumf_cavs}. This setup, assuming a volume of \SI{11.1}{\milli \liter} for each individual haloscope (obtained previously) and coherent sum, produces a total volume of \SI{0.18}{\liter}.

\subsection{Coupling system}
\label{sec:coupling_system}
The classical coupling system for extracting the energy from a resonant haloscope is a probe connected to a coaxial line. The probe is usually a monopole or a loop, depending on the type of coupling, electric or magnetic, respectively. Nevertheless, this kind of coupling is not useful at W band (\SIrange[range-phrase=--, range-units=single]{75}{110}{\giga \hertz}) due to the high attenuation levels of coaxial cables at these high frequencies. Instead, a waveguide connected to a horn antenna is used to transmit the extracted power to the receiver, as explained in the next section. In this case, the coupling element between the cavity and the waveguide is an iris which provides electrical or magnetic coupling~\cite{collins2001}, depending on the extraction position along the cavity. In this proposal rectangular irises are used, designed to obtain a critical coupling regime yielding the maximum detected power. From equation (\eqref{Pd_eq}), this is accomplished for $\beta=1$.\footnote{The factor $\kappa = \beta/(1+\beta)$ sometimes appears in the literature and corresponds to the fraction of generated power extracted from the cavity. The critical coupling regime then corresponds to $\kappa = 1/2$.}

Additionally, another port for monitoring the behavior of the cavity, the measurement of the resonant frequency and the quality factor is necessary. Unlike the detection port, this one must be highly decoupled to interfere as little as possible with the detection operation.

\subsection{Tuning system}
\label{subsec:Tuning system}
\begin{figure}[ht]
    \centering
    \includegraphics[scale=0.9]{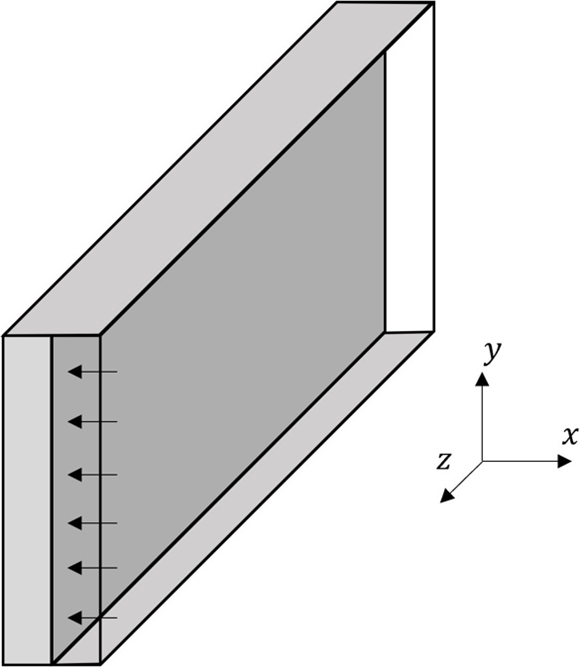}
    \caption{Tuning mechanism with sliding wall.}
    \label{fig:Tuning_mechanism_with_sliding}
\end{figure}

Exploring a wide frequency range in our experiment demands modifying the resonant frequency of each cavity. This can be achieved by modifying the cavity geometry while avoiding a high impact on operational parameters, such as the form factor, quality factor or volume. We propose to modify the cavity width, the geometry parameter that most influences the resonant frequency, by sliding a metallic wall moving along the $x$ axis, as figure \ref{fig:Tuning_mechanism_with_sliding} shows. This sliding movement is constrained by the coupling iris position and width, and yields a frequency range from \SIrange[range-units=single]{90}{102}{\giga \hertz} or a $12.5\%$ relative frequency range. The unloaded quality factor is slightly reduced from $1.45 \times 10^{4}$ to $1.38 \times 10^{4}$, whilst the volume decreases from \SIrange[range-units=single]{11.1}{9.8}{\milli \liter}.

%% file: sec5.tex
\section{Optics Design and Calibration System}
\label{sec:optics}
To optimize the sensitivity of the experiment, the extremely weak axion signal generated at the haloscope must be guided in phase with minimal losses to the detection system. For W-band, quasi-optical guiding of the signal by means of reflection at several mirrors is the most efficient method. 

\begin{figure}[ht]
    \centering
    \includegraphics[scale=0.35]{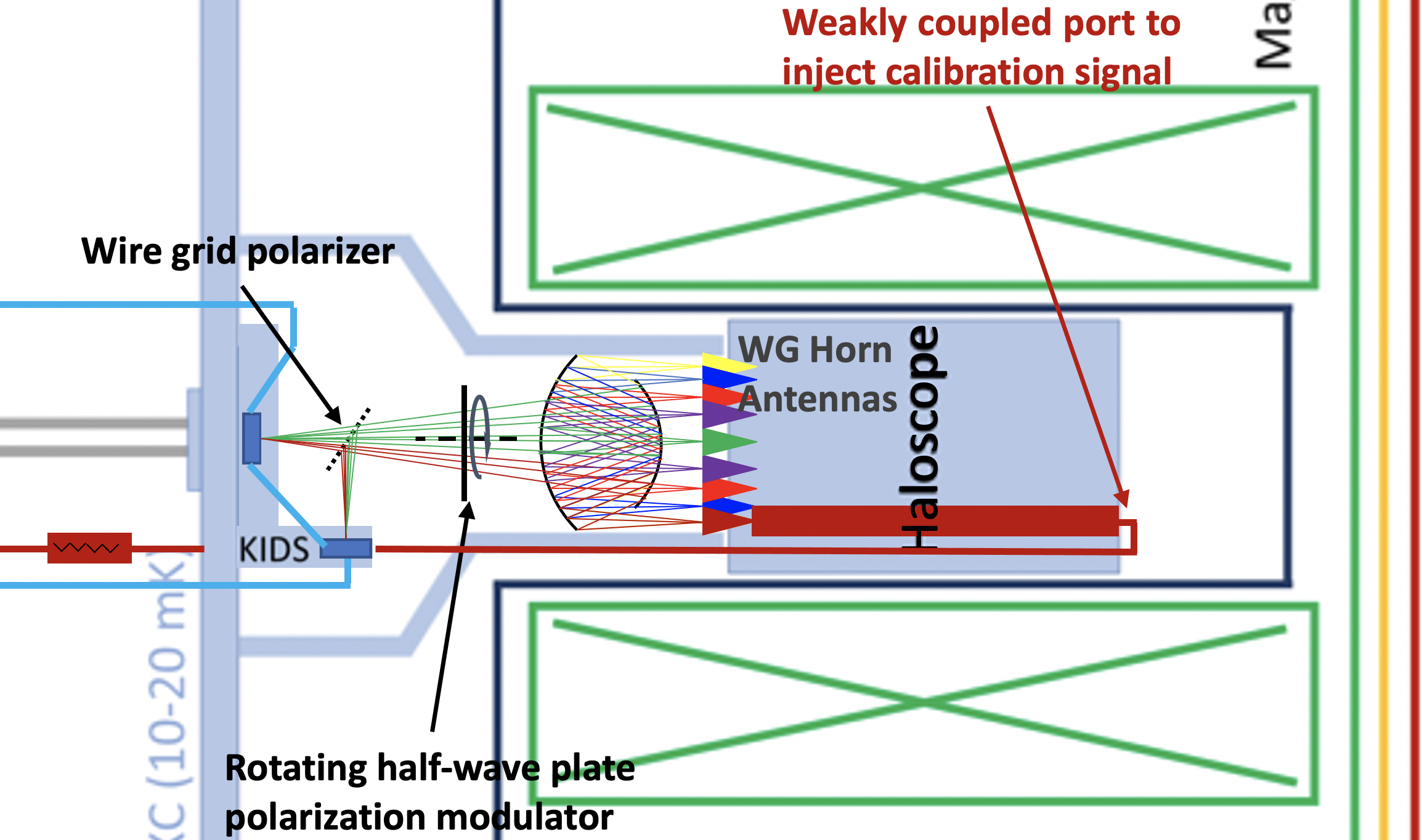}
    \caption{Zoom-in of the mK volume of figure \ref{fig:refrigerator}, showing a detailed view of the CADEX subsystems accommodation. The color scheme and subsystems are the same as in figure \ref{fig:refrigerator}. Polarized signals generated in the haloscope (coloured horns and dashed lines), are combined by an optical system (black curved vertical mirrors) and are focused on the KID arrays. Before reaching the KID detectors, the signal is modulated with a rotating half wave plate (black vertical line) and split in the two polarizations by a wire grid polarizer (tilted dotted line), allowing the full characterization of the signal's linear polarization.}
    \label{fig:model_elec_coup}
\end{figure}

Figure \ref{fig:model_elec_coup} shows the optical system design, where the signals from each individual cavity of the haloscope are collected, redirected and collimated onto the KID detectors. The optics and the haloscope are designed to guarantee the phase coherence among the cavities at the detectors, using the haloscope configuration shown in figure \ref{fig:circumf_cavs}, where all cavities are arranged in a circumference to have the same optical path from the haloscope to the detectors. In addition, the optics design minimizes phase aberration by  reducing path length differences in all mirrors, which are also large enough to minimize spillover losses.

The proposed design (see figure \ref{fig:refrigerator}) is based on a double reflector configuration fitting into the available cryostat volume, where the larger main reflector has a central hole to allow the reflected signal from the secondary smaller reflector to pass through, see figure \ref{fig:model_elec_coup}. The main reflector may be a single mirror or a set of reflectors arranged in a ring to minimize losses. The shape of the secondary reflector is designed to focus the cavity beams in the detector focal plane.

The design considers a calibration system based on the application of a Synthetic Axion Generator, similar to the one used in the ADMX experiment~\cite{ADMX_2021}. In our case, a millimeter-wave signal mimicking the one generated by the axion in a resonant cavity will be synthesized by a pulse signal generator and a high frequency analog signal generator. The signal will be injected into a resonant cavity by a weakly coupled port and calibration will be achieved by changing the output power by means of attenuators. The calibration system will also be used to test the functionality of the experiment.

%% file: sec6.tex
\section{Detection System: Kinetic Inductance Detectors}
\label{sec:detect}

The CADEx detection system will be based on state-of-the-art superconducting Kinetic Inductance Detectors (KIDs), which are high quality factor superconducting resonators indirectly coupled to a single transmission line. The working principle is based on the variation of superconducting properties caused by incoming radiation. Absorbed photons change the quasiparticle density which modifies the kinetic inductance of the resonator, lowering the resonant frequency and diminishing the quality factor of the resonator. As usual for pair-breaking detectors, the cut-off frequency that can be absorbed is limited by twice the superconducting gap, $2\Delta\approx3.52\, k_\text{B} T_\text{c}$, where $k_\text{B}$ is the Boltzmann constant and $T_\text{c}$ the superconducting critical temperature \cite{day2003broadband}. Therefore, the detection in W-band intended in the CADEx experiment requires employing a Titanium (Ti)/Aluminium (Al) bi-layer approach, which has demonstrated good sensitivity down to \SI{80}{\giga \hertz}~\cite{catalano2020sensitivity}. Optimal performance of KIDs is achieved when detectors are cooled down well below their critical temperature ($T_{\text{op}} < T_\text{c}/6$), so a cryogenic system with base temperature $\sim$ \SI{100}{\milli \kelvin} is planned. Moreover, these superconductor detectors are inherently multiplexable in the frequency domain, allowing thousands of pixels to be read-out over a single transmission line. Thus, large format KID arrays with thousands of pixels can be implemented to further increase the sensitivity. 

KIDs have been developed in the context of astronomical experiments, demonstrating state-of-the-art sensitivity ranging from the millimeter to the ultraviolet range \cite{o2019energy, NIKA2_aya}. Also, future far-infrared (FIR) missions such as the Origins Space Telescope have selected KIDs as their base technology \cite{hailey2021kinetic}. KIDs have also been proposed for dark matter experiments as indirect detectors via the absorption of athermal phonons \cite{cardani2021final, colantoni2020bullkid} or, more recently, as direct photon detectors through a broad-band haloscope
\cite{BREAD:2021tpx}.

The baseline of the detection system for CADEx aims at lumped-element KIDs (LEKIDs), where the superconducting inductor acts as the effective optical absorber of the incident radiation. To maximize the optical efficiency, the inductor geometry should be matched to the free-space impedance optimizing the meander geometry, substrate and superconducting material thicknesses and back-short distance \cite{BAja2021}.

The resonant frequency shift of the LEKIDs is read out by a single transmission line coupled to the detectors. This coupling coefficient can be tuned using low-frequency simulations by changing the separation to the line. The LEKIDs response is maximized when critical coupling is achieved under the desired optical load, or when the external quality factor ($Q_c$) equals the internal quality factor ($Q_i$). Since $Q_i$ is set by fixed parameters such as the optical background or operating base temperature, $Q_c$ will be optimized by tuning the geometrical parameters \cite{BAja2021}.

\begin{figure}[h!]
    \centering
    \includegraphics[scale=0.12]{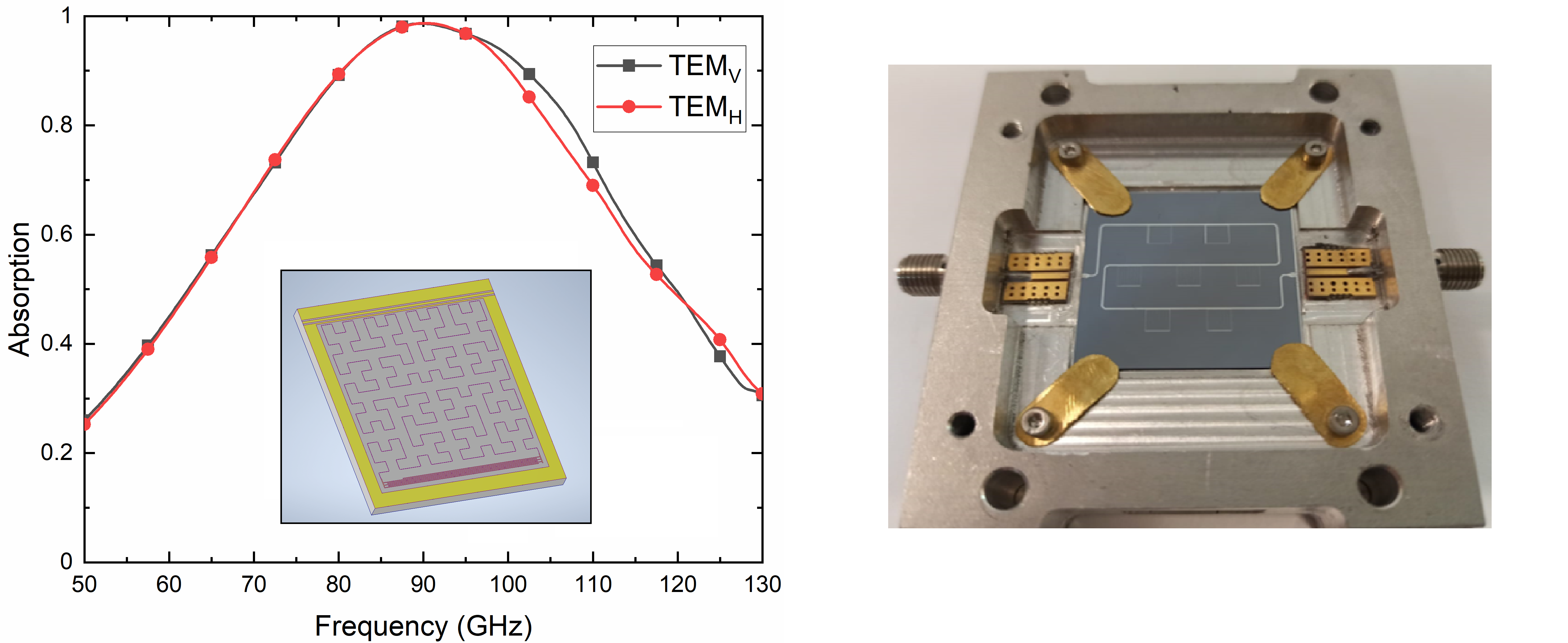}
    \caption{Left: Single LEKID using a Hilbert geometry (\SI{3}{\milli \meter} x \SI{3}{\milli \meter} cell) and absorption efficiency simulated at the W-band; Right: Seven KIDs array mounted on an aluminum holder.}
    \label{fig:Hilbert_KIDs}
\end{figure}

The CADEx experiment will search for the axion using the expected signal polarization generated in the haloscope. To measure the polarization, the detection system will follow the configuration of the polarimeters used in radio astronomy operating at frequencies above \SI{90}{\giga \hertz} like  NIKA2 at the IRAM \SI{30}{\meter} telescope \cite{NIKA2_aya, shu2018prototype}. Basically, the radiation from the haloscope will be first modulated by a half-wave polarization modulator followed by a grid polarizer which separates the two orthogonal linear polarizations to be simultaneously detected by two different LEKIDs arrays perpendicularly oriented and sharing the same read-out line. For alignment purposes, the LEKIDs will be based on a fractal Hilbert geometry with no preferential direction in absorption \cite{shu2018prototype}. Figure \ref{fig:Hilbert_KIDs} shows a preliminary single LEKID design with a Hilbert geometry and its simulated absorption for two orthogonal polarizations at W-band, as well as a visualization of an array assembled on a holder for its characterization. The proposed design allows for characterization of the polarization from the axion-photon conversion and simultaneous subtraction of the unpolarized background, for all the observing time. 

A key parameter for this experiment is the ultimate sensitivity of the detection system, the NEP, defining the weakest signal detectable by the detector. Special attention is paid to the maximum allowable magnetic field in the KIDs region, which can degrade their sensitivity. A dedicated magnetic shield and vortex traps will be developed for minimizing the effects of the magnetic field on the final sensitivity. The  sensitivity of KIDs at low radiation power is limited by the generation and recombination of quasi-particles in thermal equilibrium, which depend on superconducting properties. Visser et al.\ have already reached this limit, $\mathrm{NEP} = \SI{3.8e-19}{\watt \per \sqrt{\hertz}}$, in a low background configuration~\cite{Visser2014fluctuations}. Nevertheless, several strategies such as volume reduction and optimization of the two-level system (TLS) noise predict the potential for further improvement in  sensitivity  down to \SI{1e-20}{\watt \per \sqrt{\hertz}}, reaching the ultimate sensitivity for the proposed experiment \cite{hailey2021kinetic}.

%% file: sec7.tex
\section{Projected Axion Sensitivity}
\label{sec:sens}
In order to estimate the sensitivity of CADEx to axion-photon conversion and other signals, we compare the detected signal power $P_d$ with the expected noise in the measured power in the detector $\sigma_P$. For a haloscope using KIDs (see section~\ref{sec:detect}), the noise in the measured power depends on the NEP and scales with the exposure time $t$ as $\sigma_P = \mathrm{NEP}/\sqrt{2 t}$. 
The signal-to-noise ratio of a potential signal is then estimated as $\mathrm{SNR} = P_d/\sigma_P$.

The signal power for axion-photon conversion is given in equation \eqref{Pd_eq}. The reach of a haloscope experiment in terms of the axion-photon coupling $g_{a\gamma}$ at a desired SNR is therefore given by~\cite{Melcon:2018dba,AlvarezMelcon:2020vee}:
\begin{equation}
    g_{a\gamma}[\mathrm{GeV}^{-1}] = \left(\frac{3.88 \times 10^2 }{B[\mathrm{T}]}\right)\sqrt{\frac{(1+\beta)^2}{\beta}}\sqrt{\frac{\textrm{SNR} \,m_a [\mathrm{eV}]\,\mathrm{NEP}[\mathrm{W}/\sqrt{\mathrm{Hz}}]}{V [\mathrm{L}]\,Q_0\,t[\mathrm{s}]^\frac{1}{2}\, C}}\,.
    \label{eq:HaloScope_Reach}
\end{equation}
Here, we have assumed an axion density of $\rho_a = \SI{0.45}{\giga \eV \per \cubic \centi \meter}$~\cite{Read:2014qva}. We will consider a baseline configuration for CADEx with a cavity volume of \SI{0.2}{\liter}, a magnetic field of \SI{10}{\tesla} and a cavity quality factor of $Q_0 = 2 \times 10^{4}$. We fix $\beta = 1$, the optimal coupling between the cavity and receiver to extract the maximum power. Following the discussion in section \ref{sec:form_factor}, we set $C \approx 0.66$. The axion mass $m_a$ needs to match the resonant frequency of the cavity.

\begin{figure}[tb]
    \centering
    \includegraphics[width=0.95\textwidth]{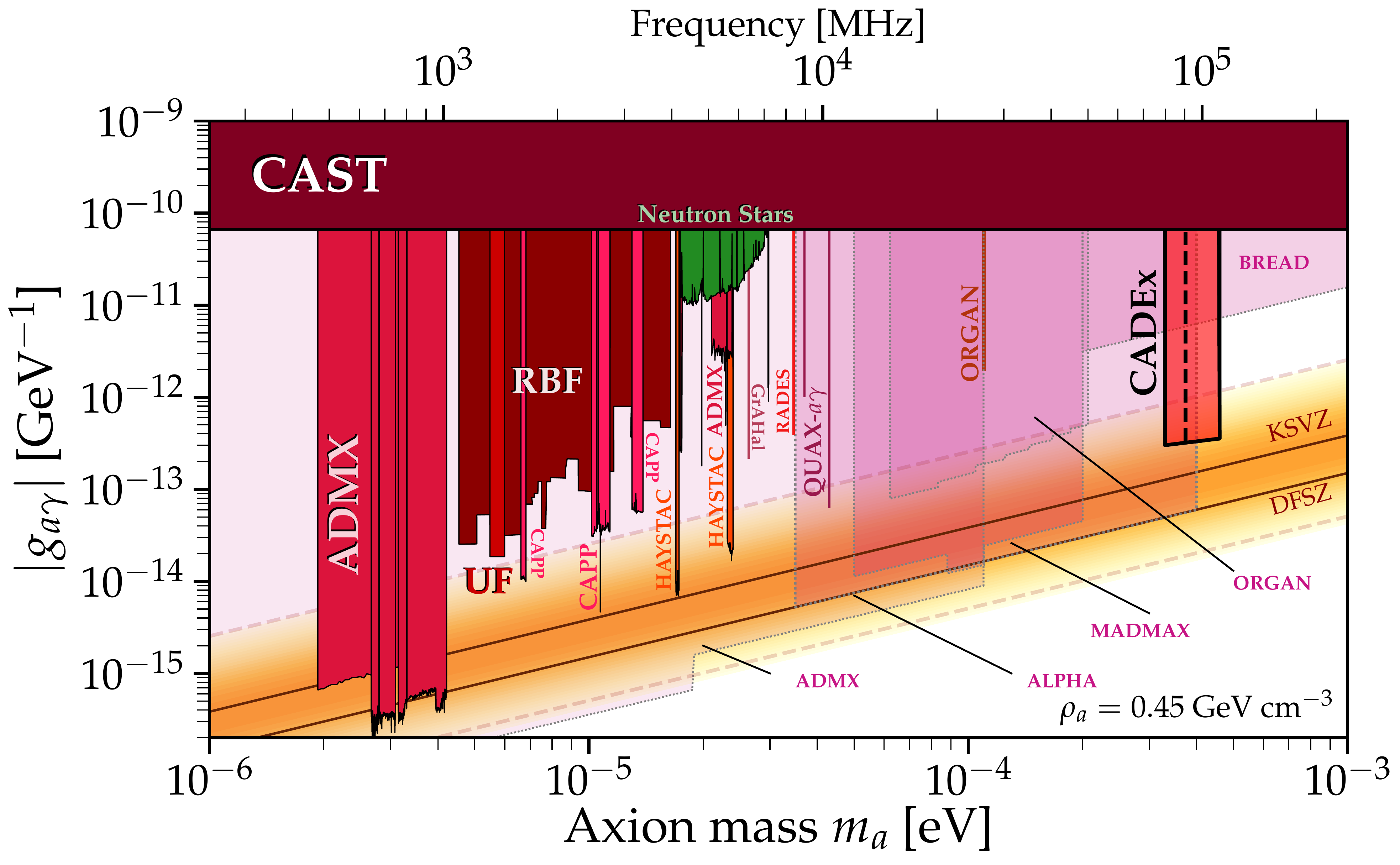}
    \caption{\textbf{Projected CADEx sensitivity to the axion-photon coupling $g_{a\gamma}$.} The vertical black dashed line corresponds to the $5\sigma$ sensitivity ($\mathrm{SNR} = 5$) of a 1-year exposure with noise equivalent power $\mathrm{NEP} = \SI{1e-19}{\watt \per \sqrt{\hertz}}$. The region bounded by a solid black line corresponds to the sensitivity with roughly 3000 4-day exposures and \SI{1e-20}{\watt \per \sqrt{\hertz}}, achievable on a timescale of $\mathcal{O}(30)$ years. For comparison, we show a number of existing constraints from the CAST helioscope~\cite{CAST:2017uph}, various axion haloscopes (filled red and purple regions)~\cite{DePanfilis,Hagmann,McAllister:2017lkb,Du:2018uak,ADMX:2018ogs,HAYSTAC:2018rwy,Braine:2019fqb,Alesini:2020vny,CAST:2020rlf,CAPP:2020utb,Choi:2020wyr,Zhong:2018rsr,Backes:2020ajv,Grenet:2021vbb,ADMX:2021nhd}, and neutron stars~\cite{Foster:2022fxn}, along with projected constraints from other proposed haloscopes (transparent red regions)~\cite{McAllister:2017lkb,Lawson:2019brd,Beurthey:2020yuq}. Figure adapted from~\cite{ohare}.} 
    \label{fig:AxionSensitivity}
\end{figure}

The projected $5\sigma$ sensitivity of CADEx is shown in figure \ref{fig:AxionSensitivity}. The vertical black dashed line corresponds to a one year search
centered on an axion mass of $m_a = \SI{372}{\micro \eV}$, assuming  $\mathrm{NEP} \approx \SI{1e-19}{\watt \per \sqrt{\hertz}}$, based on presently available KIDs technology. This search would achieve a sensitivity down to $g_{a\gamma} \approx \SI{3e-13}{\per \giga \eV}$.

Exploring a wider range of axion masses requires a large number of searches with the haloscope tuned to different resonant frequencies $\nu_c$. The cavity bandwidth is $\Delta\nu_c = \nu_c/Q_\ell \approx \SI{9}{\mega \hertz}$. A frequency range of \SI{30}{\giga \hertz} (corresponding to axion masses \SIrange[range-phrase=--, range-units=single]{330}{460}{\micro \eV}) could be covered with $\sim 3000$ exposures. Assuming that a NEP of \SI{1e-20}{\watt \per \sqrt{\hertz}} can be achieved with future technology, sensitivity down to $g_{a\gamma} \approx \SI{3e-13}{\per \giga \eV}$ can be achieved in a single exposure of $\sim 4$ days. The mass range \SIrange[range-phrase=--, range-units=single]{330}{460}{\micro \eV} could therefore be probed on a total measuring time of $\sim$ 30 years (which can be split among several instruments with haloscopes build for different frequency ranges), shown by the region bounded by a thick black solid line in figure \ref{fig:AxionSensitivity}.

For comparison, we also show in figure~\ref{fig:AxionSensitivity} a number of existing constraints from axion haloscopes, from neutron stars observations and from the CAST helioscope. These provide constraints on the axion parameter space for $g_{a\gamma} \gtrsim \SI{7e-11}{\per \giga \eV}$ and for $m_a \lesssim \SI{50}{\micro \eV}$. CADEx would probe unexplored parameter space at higher masses, well-motivated by cosmological production mechanisms, reaching into the region of axion-photon couplings suggested by QCD axion theory~\cite{DiLuzio:2016sbl,DiLuzio:2017pfr,Agrawal:2017cmd}, shown by the yellow band in figure~\ref{fig:AxionSensitivity}. CADEx would be complementary to other proposals using established search techniques with resonant cavities, such as ADMX~\cite{Stern:2016bbw} and ORGAN~\cite{McAllister:2017lkb}, which should have sensitivities up to masses of \SI{200}{\micro \eV}, as well as alternative broadband detector concepts such as ALPHA, MADMAX and BREAD.\footnote{We plot projections for BREAD assuming 1000 days of exposure and baseline assumptions on NEP from reference~\cite{BREAD:2021tpx}.}

An important advantage of our KIDs detection system is that 
modulating the 
haloscope signal as a function of 
polarization allows for distinguishing the axion signal from background unpolarized systematics. A true axion signal is detected as an excess of power in one of the 
frequency channels scanned by the haloscope over the neighboring ones, which appears only in the polarization expected for the axion. The proportionality of the signal to $B^2$ can also be tested.

\subsection{Dark photon sensitivity}
Constraints on dark photons can be derived similarly to constraints on axions and axion-like particles. The signal power due to the resonant conversion of dark photons can be obtained from equation \eqref{Pd_eq}, using the correspondence~\cite{Arias:2012az,Ghosh:2021ard}:
\begin{equation}
    g_{a\gamma} \rightarrow \frac{\chi m_{\gamma^\prime} \sqrt{ \cos^2\theta_\mathrm{pol}}}{B}\,,
    \label{eq:dark_photon_map}
\end{equation}
where we assume that the dark photons account for all of the local DM. Here, $\chi$ is the kinetic mixing parameter and $\theta_\mathrm{pol}$ is the angle between the polarization vector of the dark photon field and the electric field polarization to which the detector is sensitive. Using the detection system described in section~\ref{sec:detect}, we propose to measure the two orthogonal linear polarizations of the photon signal. In this case, $\theta_\mathrm{pol}$ is the angle between the dark photon polarization and the plane defined by these orthogonal photon polarizations. 

Depending on the cosmological production and evolution of the dark photon field (e.g.~\cite{Co:2018lka,Bastero-Gil:2018uel,Agrawal:2018vin,Co:2021rhi}), the dark photon may be polarized along a fixed direction on long timescales compared to the integration time of the experiment at a given frequency channel. In this case, the dark photon polarization remains fixed but $\theta_\mathrm{pol}$ varies with time as the orientation of the detector changes with the Earth's rotation. This gives rise to a periodic signal $P(t) \propto \cos^2\theta_\mathrm{pol}(t)$ with a period of 1 day and an $\mathcal{O}(1)$ oscillation amplitude~\cite{Caputo:2021eaa}. In this scenario, the detection of a time-varying polarized signal could be used to discriminate from backgrounds and claim a discovery of the dark photon, as long as the coherence time for the polarization exceeds $\sim 4$ days (the duration of each mass scan). The value of $\theta_\mathrm{pol}$ averaged over long timescales depends on the detector orientation and the dark photon polarization direction. We therefore fix $\langle\cos^2\theta_\mathrm{pol} \rangle = 2/3$, the average over randomly oriented dark photon polarization angles~\cite{Caputo:2021eaa}.\footnote{This factor of 2/3 can be intuitively understood from the fact that at any given time, the detector will be sensitive to two of the three possible polarization directions of the dark photon. However, we need to take into account that the plane of polarization of the converted photons is not known a priori, implying that various orientations of the splitter would need to be tested at each scanned frequency.} Under these assumptions, it is possible to map the projected sensitivities in the axion parameter space $(m_a, g_{a\gamma})$ to the dark photon reach in $(m_{\gamma^\prime}, \chi)$, using equation~\eqref{eq:dark_photon_map}.

\begin{figure}[tb!]
    \centering
    \includegraphics[width=0.95\textwidth]{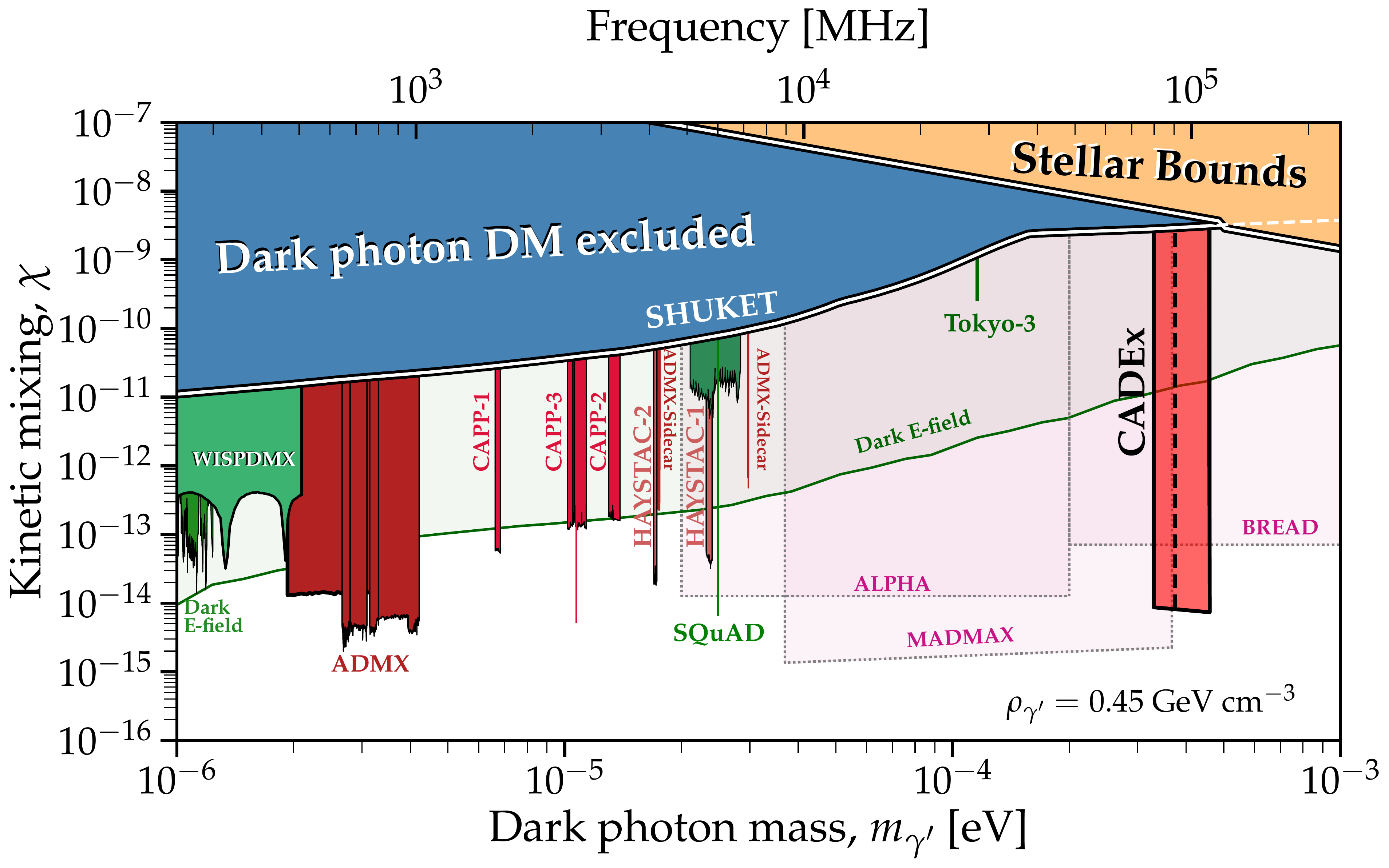}
    \caption{\textbf{Projected CADEx sensitivity to the dark photon kinetic mixing $\chi$.} The region labelled ``CADEx'' is the sensitivity achievable using the same searches as presented in figure~\ref{fig:AxionSensitivity} (with no additional data taking). The solid blue regions show where dark photons are excluded from being all of the Dark Matter on cosmological grounds~\cite{Arias:2012az,Caputo:2021eaa}. The orange region shows the envelope of constraints from stellar cooling (see \cite{Caputo:2021eaa} for a compilation). We show current and projected constraints from other axion haloscopes in red, with dedicated dark photon searches shown in green~\cite{Brun:2019kak,Nguyen:2019xuh,Tomita:2020usq,Dixit:2020ymh,Godfrey:2021tvs}.  Figure adapted from \cite{Caputo:2021eaa} and \cite{ohare}.} 
    \label{fig:DarkPhotonSensitivity}
\end{figure}

In figure \ref{fig:DarkPhotonSensitivity}, we show the 5$\sigma$ sensitivity to the dark photon kinetic mixing which can be achieved by CADEx in this \textit{constant, fully polarized} dark photon scenario, using the same data taken for the axion search described above and presented in figure \ref{fig:AxionSensitivity}.  CADEx should be sensitive to values of $\chi \sim  10^{-14}$ for dark photon masses in the range \SIrange[range-phrase=--, range-units=single]{330}{460}{\micro \eV}. Existing constraints in this region are at the level of $\chi \sim 10^{-9}$, where cosmological constraints~\cite{Arias:2012az,Caputo:2021eaa} and stellar cooling constraints~\cite{Caputo:2021eaa} intersect. CADEx will therefore significantly enhance sensitivity in this region of dark photon parameter space. 

We have so far considered the \textit{fixed, full polarization} scenario for the dark photon. However, the evolution of the dark photon through structure formation may wash out any initial large-scale polarization which may be present. In an \textit{unpolarized} scenario, the resulting photon signal would also be unpolarized, with no time-variation to distinguish it from backgrounds. Moreover, the dark photon signal does not rely on the presence of the magnetic field, meaning that it is not easy to obtain a `background-only' data set for comparison. In principle, it may still be possible to set limits on $\chi$, by searching for a change in the intensity measured by the KID detector at a frequency channel compared to neighboring ones. However, this would require a good control of systematics in the absolute power calibration of the KIDs detectors and in any background radiation in the experiment over a range of frequencies. In the \textit{unpolarized} scenario, then, figure~\ref{fig:DarkPhotonSensitivity} would represent the most optimistic possibility for setting limits on $\chi$, assuming that absolute background systematics can be corrected. However, we emphasize that this is likely to be experimentally challenging and that a more detailed understanding of the expected polarization of the dark photon in the Milky Way will be essential to characterize this signal in the future.


%% file: sec8.tex
\section{Conclusions}
\label{sec:con}
The QCD axion arises naturally as an extension of the Standard Model to solve the strong CP problem. Simultaneously, axions may be produced with the correct abundance in the early Universe to provide the current cold dark matter content. Haloscopes are being widely used to search for the QCD axion in the mass range \SIrange[range-phrase=--, range-units=single]{1.65}{49.6}{\micro \eV}. However, searches for the axion above this mass range, well motivated by theory, have not yet been performed primarily due to a number of technological challenges in haloscopes and in the detection system. Overcoming these challenges will allow sensitive searches for axions in the dark matter halo with masses 
in the range \SIrange[range-phrase=--, range-units=single]{100}{1000}{\micro \eV}, as well as other light new particles such as the dark photon. 

This paper presents CADEx, a novel experiment to search for the Dark Matter QCD axion and dark photon in the unexplored mass range \SIrange[range-phrase=--, range-units=single]{330}{460}{\micro \eV} operating within the W-band at the Canfranc Underground Laboratory (Spain). CADEx will push the microwave resonant cavity haloscope technology to high frequencies, increasing its collecting power by means of the coherent sum of multiple large cavities. The detection system uses the  polarization properties of the axion signal arising from the haloscope, 
and is based on broadband Kinetic Inductor Detectors (KIDs) with sensitivities that have a strong improvement potential in the near future. When equipped with a \SI{0.2}{\liter} haloscope in a high and static magnetic field of \SIrange[range-phrase=--, range-units=single]{8}{10}{\tesla} and a detection system with  KIDs sensitivities of \SI{1e-20}{\watt \per \sqrt {\hertz}}, CADEx will provide a sensitivity three orders of magnitude better than the current best limit from CAST~\cite{CAST:2020rlf}, reaching the well-motivated region for  QCD axion dark matter predicted from models~\cite{DiLuzio:2016sbl,DiLuzio:2017pfr,Agrawal:2017cmd}. 

The same setup will also provide sensitivity down to a dark photon kinetic mixing of  $\chi \sim  10^{-14}$ over the same range of masses, for the case where the dark photon is fully polarized. 
This would provide  around 5 orders of magnitude improvement over current constraints, in a region of parameter space where the dark photon may account for all of the dark matter~\cite{Arias:2012az} while constraints from stellar cooling are weak~\cite{Caputo:2021eaa}. Although it is not clear at present if these constraints can still be obtained when the dark photon is unpolarized, we note that the dark photon search can be performed with data obtained
when the magnetic field is turned off for the purpose of testing the background systematics that may also affect the axion search.

CADEx will  provide a multidisciplinary platform to develop novel concepts of haloscopes, including  tunable-cavity haloscopes, and push the W-band superconducting detectors to their ultimate sensitivities,
which are crucial to confirm or rule out the QCD axion predicted by the models with a mass in the range \SIrange[range-phrase=--, range-units=single]{330}{460}{\micro \eV}.